# Interdisciplinarity as Diversity in Citation Patterns among Journals: Rao-Stirling Diversity, Relative Variety, and the Gini coefficient

Loet Leydesdorff,*[a] Caroline S. Wagner,[b] and Lutz Bornmann[c]

**Abstract**

Questions of definition and measurement continue to constrain a consensus on the measurement of interdisciplinarity. Using Rao-Stirling (RS) Diversity produces sometimes anomalous results. We argue that these unexpected outcomes can be related to the use of "dual-concept diversity" which combines "variety" and "balance" in the definitions (*ex ante*). We propose to modify RS Diversity into a new indicator (DIV) which operationalizes "variety," "balance," and "disparity" independently and then combines them *ex post*. "Balance" can be measured using the Gini coefficient. We apply DIV to the aggregated citation patterns of 11,487 journals covered by the Journal Citation Reports 2016 of the *Science Citation Index* and the *Social Sciences Citation Index* as an empirical domain and, in more detail, to the citation patterns of 85 journals assigned to the Web-of-Science category "information science & library science" in both the cited and citing directions. We compare the results of the indicators and show that DIV provides improved results in terms of distinguishing between interdisciplinary knowledge integration (citing) versus knowledge diffusion (cited). The new diversity indicator and RS diversity measure different features. A routine for the measurement of the various operationalizations of diversity (in any data matrix) is made available online.

[a] *corresponding author; Amsterdam School of Communication Research (ASCoR), University of Amsterdam
PO Box 15793, 1001 NG Amsterdam, The Netherlands; loet@leydesdorff.net
[b] John Glenn College of Public Affairs, The Ohio State University, Columbus, Ohio, USA, 43210; wagner.911@osu.edu
[c] Division for Science and Innovation Studies, Administrative Headquarters of the Max Planck Society, Hofgartenstr. 8, 80539 Munich, Germany; bornmann@gv.mpg.de

## 1. Introduction

Policymakers and researchers continue to be interested in measures of "interdisciplinarity" (Wagner *et al*., 2011). Recently, a great deal of attention has been paid to using references as a way to measure "interdisciplinarity" (e.g., Boyack & Klavans, 2014; Mishra & Torvik, 2016; Tahamtan & Bornmann, 2018; Wang, 2016), These analyses are notable because of the increasing consensus, following Rao (1982) and Stirling (2007), for defining interdisciplinarity as diversity encompassing three features: variety, balance, and disparity. However, a problem arises when measuring the interrelationships among the three components: how can they be combined without losing either information or validity?

In this study, we revisit the definition and measurement of variety, balance, and disparity; we compare methods; and we elaborate an approach that addresses the anomalous outcomes that can occur when using Rao-Stirling (RS) Diversity. Leydesdorff *et al.* (2017), for example, compared twenty cities in terms of the RS diversity of patent portfolios; the results showed the unsatisfying finding that Rotterdam and Jerusalem scored above Shanghai and Paris. The new measure applied to this same data reverses the order: Shanghai is ranked in the first place, and Rotterdam only in the 16$^{th}$ place among twenty cities. At that time, the new method of measurement was proposed, but without substantial empirical testing.

In RS diversity, two of the three components (variety and balance) are combined in the definitions (*ex ante*) using the Simpson Index. However, these components can be measured *independently* and thereafter combined (*ex post*). We argue that the *ex ante* definition of "dual-concept" diversity (e.g., the Simpson Index) is already known to be a source of unnecessary distortions (Stirling, 1998, pp. 48f.). Using the Gini coefficient, however, "balance" can be operationalized independently (Nijssen, Rousseau, & Van Hecked, 1998), as can "variety."

We discuss these different measures applied to interdisciplinarity in terms of diversity and compare the empirical results. To this purpose, we use the full set of 11,487 journals contained in the Journal Citations Report 2016 of the *Science Citation Index* and the *Social Sciences Citation Index,* with the question whether and to what extent the different indicators measure the same or different dimensions of or perspectives on "interdisciplinarity." A case study using the aggregated citations among the 85 journals assigned within the larger set to the Web-of-Science Subject Category (WC) for "information science & library science" is further elaborated. Does the new diversity measure improve on the measurement of interdisciplinarity in comparison to RS diversity?

A routine for the computation is provided at https://leydesdorff.github.io/diversity_measurement/ (and http://www.leydesdorff.net/software/mode2div). The routine can be used to compute RS diversity, the new diversity measure DIV, and the respective components in any data matrix (e.g., a word/document matrix or a citation matrix) written as a Pajek file.[4] In this study, we chose the matrix of 10,000+ journals contained in the Journal Citation Reports 2016 citing each other. This provides a large empirical domain with which we are familiar from previous studies and in which we encountered the problems using RS diversity for the measurement of interdisciplinarity (Leydesdorff, Wagner, & Bornmann, 2018). However, the measure can be applied to any data matrix (e.g., Bache *et al*., 2013).

---

[4] The Pajek format can be used for virtually unlimited large matrices and is readily available in most network analysis and visualization programs. UCInet offers the option to rewrite Excel files in the Pajek format.

## 2. The Measurement of Interdisciplinarity in terms of Diversity

In a paper, ambitiously entitled "A general framework for analysing diversity in science, technology and society," Stirling (2007) addressed the problem of measuring interdisciplinarity based on his extensive review of the methodological and statistical literature (Stirling, 1998). Stirling's (2007) study has been influential in science & technology studies, to the extent that Rafols & Meyer (2010, pp. 266f.) have defined the "Stirling Index" of diversity—with only a footnote mentioning Rao's (1982) original formulation—as follows:[5]

$$\Delta = \sum_{i,j} d_{ij}(p_i p_j) \qquad (1)$$

In this equation, $p_i$ is the proportion of elements assigned to each class *i* and $d_{ij}$ denotes a disparity measure between the two classes *i* and *j*. Note that the classes can be defined at different levels of aggregation. For example, one can measure the diversity of references in articles in terms of the cited journals or in terms of the WCs attributed to the journals. The resulting values of *Δ* will be different. Analogously, diversity will vary with the number of digits in the case of using Medical Subject Heading (MeSH) or patent classes.

---

[5] Stirling (2007, at p. 712) formulated the most general case of $\Delta = \sum_{i,j}(p_i p_j)^\alpha d_{ij}^\beta$. The introduction of exponents opens another parameter space. In most scientometric applications, authors assume the reference case of α = β = 1.

In the scientometric literature, this measure Δ is widely called the "Rao-Stirling diversity indicator" (e.g., Cassi, Champeimont, Mescheba, & de Turckheim, 2017) as different, for example, from Simpson diversity (Simpson, 1949) or Shannon entropy (Shannon, 1948). The right-most factor of Eq. 1 [$\sum_{i,j}(p_i p_j)$] is also known as the Hirschman-Herfindahl Index in economics or the Simpson index in biology.[6] The first term of the equation [$\sum_{i,j} d_{ij}$] adds the distribution in a (e.g., geographical) space. For example: if distances in a subset are small, this space can be considered as a niche of "related variety" (Frenken, Van Oort, & Verburg, 2007).

**Table 1** Selected measures of diversity

| Notation | |
|---|---|
| Proportion of elements in category $i$: | $p_i$ |
| Distance between categories $i$ and $j$: | $d_{ij}$ |
| Similarity between categories $i$ and $j$: | $s_{ij} = 1 - d_{ij}$ |
| **Indices** | |
| $N$ = Variety | $N$ |
| $H$ = Shannon | $-\sum_i p_i \ln p_i$ |
| $I$ = Simpson diversity | $\sum_{i,j(i \neq j)} p_i p_j = 1 - \sum_i p_i^2$ |
| $\Delta$ = Stirling ($\alpha = 1$, $\beta = 1$) | $\sum_{i,j} d_{ij} p_i p_j = 1 - \sum_{i,j} s_{ij} p_i p_j$ |
| Generalised Stirling | $\sum_{i,j} d_{ij}^\alpha (p_i p_j)^\beta$ |

Table 1 [from Rafols & Meyer (2010, p. 267), but based on Stirling (2007, p. 709)] summarizes the distinctions among the various indicators of diversity. However, we agree with Stirling's crucial argument that diversity—and by implication "interdisciplinarity"—is composed of *three* (and not more than three) components which he labeled "variety," "balance," and "disparity." He formulated the relations among these three components as follows:

---

[6] $\Sigma_{ij} p_i p_j = 1$ when taken over all $i$ and $j$. The Simpson index is equal to $\Sigma_i (p_i)^2$, and the Gini-Simpson to $[1 - \Sigma_i (p_i)^2]$. See also Table 1.

Each is a necessary but insufficient property of diversity (Sokal & Sneath, 1970; Clarke, 1978, Stirling, 2006d). Although addressed in different vocabularies, each is applicable across a range of disciplines and aggregated in various permutations in quantitative indices (Hill, 1973). Despite the multiple disciplines and divergent contexts, there seems no other obvious candidate for a fourth important general property of diversity beyond these three (Stirling, 2006e).

In other words: all else being equal, for each of these three factors (that is, *ceteris paribus*), an increase in one of the three components leads always to a greater diversity. This has also been called "the monotonicity" requirement (Rousseau, 2018a): diversity increases for each of the three components when the other two remain the same.

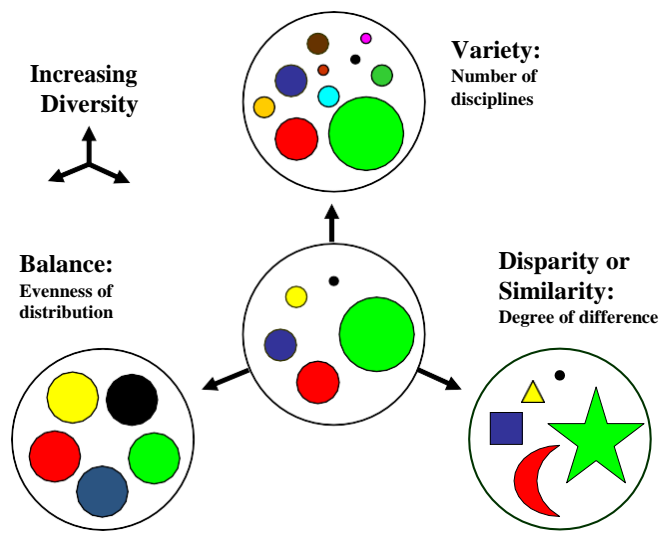

**Figure 1:** Schematic representation of the attributes of diversity, based on Stirling (1998, p. 41). Source: Rafols & Meyer (2010, p. 266).

Rafols & Meyer (2010, p. 266) provided Figure 1, which has become iconographic for visualizing the distinctions among the three components. From the perspective of hindsight, previous attempts to operationalize "interdisciplinarity" can be recognized as using one or two of the three components of diversity suggested in Figure 1. For example, Porter & Chubin (1985) proposed using the proportion of references to sources outside the WC of the paper under study as a measure of interdisciplinary knowledge integration into the citing paper (cf. Morillo, Bordons, & Gómez, 2001; Uzzi, Mukherjee, Stringer, & Jones, 2013).[7] The focus in these studies is limited to variety. Rafols, Porter, & Leydesdorff (2010) generalized this concept of diversity into spreads in portfolios—e.g., of references across WCs—projected on a map. The map provides distances among the nodes ($d_{ij}$) that can be used for the measurement of disparity.[8]

From the perspective of network analysis, we have explored Betweenness Centrality (BC) as an indicator of diversity and interdisciplinarity (Leydesdorff, Goldstone, & Schank, 2008). Using the aggregated journal-journal citation relations provided by the Journal Citation Reports 2015 as a comprehensive set ($n > 11,000$ journals), Leydesdorff, Wagner, & Bornmann (2018) tested RS Diversity and BC against each other as measures of interdisciplinarity. However, the results were disappointing: whereas BC was found to indicate "multidisciplinarity" more than "interdisciplinarity," the authors cautioned (at p. 588) that "[…] Rao-Stirling 'diversity' is often used as an indicator of interdisciplinarity; but it remains only an indicator of diversity." Furthermore, "the interpretation of diversity as interdisciplinarity remains the problem" and the authors warn that "policy analysts seeking measures to assess interdisciplinarity can be advised to specify first the relevant contexts […]. The arguments provided in this study may be helpful […] by specifying methodological limitations" (p. 589). As one of the possible limitations, the authors mentioned the operationalization of Stirling's *three* dimensions in terms of only *two* factors being multiplied: disparity [$\sum_{i,j} d_{ij}$] and the Simpson Index [$\sum_{i,j}(p_i p_j)$].

---

[7] Analogously, one can consider diversity in the cited direction as knowledge diffusion (Rafols, 2014; cf. Rousseau, Zhang, & Hu (in preparation).
[8] The map provides a two-dimensional projection of an n-dimensional space and therefore a distance on the map provides a poor indicator of distance.

In ecology, efforts have been made to integrate the two components of "variation" and "balance" (*ex ante*) into a single indicator such as the Simpson Index. This has also been called "dual concept diversity" (e.g., Junge, 1994). According to Stirling (1998, p. 48) "'dual concept diversity' has become synonymous with diversity itself to many authorities in ecology." In scientometrics, Rousseau *et al.* (1999), for example, formulated in a similar spirit as follows (at p. 213):

> It is generally agreed that diversity combines two aspects: species richness and evenness. Disagreement arises at how these two aspects should be combined, and how to measure this combination, which is then called "diversity."

Although Stirling (1998, p. 57) concluded that there are good reasons to prefer the Shannon measure above the Simpson Index *if* one wishes to measure the two concepts in a single operationalization as a "dual concept," he himself eventually chose to extend the Simpson Index—as a dual concept indicator—with disparity as a third dimension. The problem of the duality of the Simpson Index was thereby inherited into the RS diversity indicator. Stirling (1998, at p. 48) was aware of this problem when he formulated the following empirical question:

> Where a system displays simultaneously greater variety and balance, there is little need for a single integrated concept to recognise that it is intrinsically more (dual concept)[9] diverse. However, it is much more likely to be the case that no single system can be considered unequivocally to be intrinsically more diverse than others in this sense. In such cases, the crucial questions concern the relative importance assigned to variety and balance in arriving at the overall notion of diversity.

Rousseau (2018a, at p. 651) concluded in a further reflection that "the balance aspect is not hidden in the 'dual concept,' but simply is not present" in the RS measure. By providing a counter-example, the author showed that RS diversity does not meet the *ceteris paribus* monotonicity requirement which states that for a given variety and disparity, the diversity increases monotonically with balance. Rousseau added that this same conclusion—the absence of an indicator of balance—holds equally for the "true diversity" variant of RS diversity offered by Zhang, Rousseau, & Glänzel (2016). Following Leinster and Cobbold (2012), however, Rousseau (2018a and b) argues that balance is not an essential component of diversity.

In a brief communication, Leydesdorff (2018) responded to Rousseau (2018a) that there is no need for such a drastic revision of Stirling's *theoretical* conceptualization in terms of "balance," "disparity," and "variety." The problem is the operationalization: instead of *ex ante* combining "balance" and "variety," however, Nijssen, Rousseau, & Van Hecke (1998) offered a possibility of distinguishing analytically between balance and variety. They proved mathematically that both the Gini index and the coefficient of variation (that is, the standard deviation divided by the mean of the distribution or, in formula format, $\sigma/\mu$) are ideal indicators of balance. (Unlike the

---

[9] In other words, assuming that both systems display equal disparity.

Gini coefficient, however, the coefficient of variation is not bounded between zero and one.) Furthermore, the Gini index is **not** a measure of variety (Rousseau, 2018a, p. 649). In principle this conclusion enables us to distinguish operationally between "variety" and "balance" as two independent dimensions—represented by two different equations. The empirical results can then be combined by multiplying the values between zero and one *ex post.*

"Variety" can be independently operationalized—as in the number of classes ($n_c$) in use—or as relative variety (bounded between zero and one) as $n_c/N$—with $N$ being the total number of classes available. As noted, "balance" can be operationalized using the Gini coefficient without co-mingling it with "variety" (Nijssen *et al.*, 1998). Since the Gini coefficient is maximally diverse for Gini = 0 and fully homogeneous for Gini = 1, we use $(1 - \text{Gini})^{10}$ so that one obtains a diversity measure with three components for each unit of analysis *c,* as follows:

$$Div_c = (n_c/N) * (1 - \text{Gini}_c) * \sum_{\substack{i=1,\\j=1,\\i\ne j}}^{\substack{i=n_c\\j=n_c}} d_{ij} / [n_c * (n_c - 1)] \qquad (2)$$

The right-most factor in this equation is similar to (*i*) the disparity measure used in the case of RS diversity. The *two* other factors, however, represent (*ii*) relative variety as $n_c/N$—with $N$ being the total number of classes available—and (*iii*) balance measured as (1 – the Gini coefficient) of the same distribution. (Variety and disparity have to be normalized so that all terms are bounded between zero and one.)

---

[10] This is a change in the definition when compared with Leydesdorff (2018), where we used the Gini-value itself.

In Eq. 2, $n_c$ is the number of classes with values larger than zero and $N$ is the number of available classes in the domain. For example, *Scientometrics* was cited by articles in 38 of the 86 journals belonging to the WC of "information science and library science" in 2015, leading to a *relative* variety in this citation distribution of 38/86 = 0.442. In cases where the number of classes is not known, one can normalize pragmatically by using the maximum number of observed classes or, in other words, the longest vector in the reference set.

As noted, Leydesdorff (2018) compared the new measure to the RS diversity of the patent portfolios of 20 cities. However, in order to assess the quality of the two measures as an indicator of "interdisciplinarity," we required a larger data set. In this study, we return to the JCR data we used in the previous study and which resulted in unsatisfactory values for RS diversity. Does Eq. 2 provide us with more convincing results? In addition to the interpretability of the results, we can consider the (rank-order) correlation with BC across the distribution of 10,000+ journals as another indicator of the validity of the measurement. Are these indicators— RS diversity and our new measure (DIV)—significantly different in their relations to BC? Does the exercise bring us further towards indicating interdisciplinarity?

Before turning to these empirical questions, let us first consider the concept of "coherence." Rafols & Meyer (2010, pp. 268 ff.) conceptualized interdisciplinarity in terms of both diversity and coherence. Analogously, others use the words "novelty" and "conventionality" (Uzzi *et al.*, 2013; Schilling & Green, 2015; Stephan *et al.*, 2017), or 'atypical' combinations, but which are limited in terms of accounting for balance. Leydesdorff & Rafols (2011, at p. 852) and Rafols,

Leydesdorff, O'Hare, Nightingale, & Stirling (2012, at p. 1268) have proposed to operationalize coherence as follows:

$$C = \sum_{ij(i \neq j)} p_{ij} \cdot d_{ij} \qquad (3)$$

This measure of coherence accounts for both the probability of co-occurrences of classes *i* and *j* ($p_{ij}$) and the distances ($d_{ij}$) between these classes. In other words, *C* measures the average distance among classes related in a network. Coherence *C* and RS diversity can also be compared as observed ($p_{ij}$) versus expected values ($p_i * p_j$) of interdisciplinarity (Rafols, 2014).

However, it is less clear whether and how coherence scores should be combined with diversity into a composed indicator of interdisciplinarity. Diversity in the referencing can also be considered as "interdisciplinary" knowledge integration, whereas diversity in being cited has been considered as diffusion (Rousseau, Zhang, & Hu, forthcoming; cf. Leydesdorff & Rafols, 2011a and b). For the purpose of this study, we limit ourselves to the debate about diversity.

## 3. Data and Methods

*3.1. Data*

We test the measures both in the full set of the journals included in the JCRs 2016 and in the case of the subset of 85 journals subsumed by ISI/Clarivate under the WC of "information science and library science." (We used the analogous sets for 2015 in our previous study.) As being active practitioners in this field (LIS), we may be able to provide the results for the subset a more informed interpretation. Actually, this focus led to our worry about RS diversity as an indicator of "interdisciplinarity." We formulated (at pp. 579f.):

> In terms of knowledge integration indicated as diversity in the citing dimension, *JASIST* assumes the third position and *Scientometrics* trails in 45th position. In the cited dimension, the diversity of *Scientometrics* is ranked 70 (among 86). Thus, the journal [*Scientometrics*] is cited in this environment much more specifically than in the larger context of all the journals included in the JCR, where it assumed the 339th and 6,246th position among 11,359 observations, respectively.

The 70th position of *Scientometrics* within this set of 86 LIS journals is very counter-intuitive.

The new routine adds both RS diversity and the new diversity measure to a spreadsheet, as well as the other relevant indices such as Gini, Simpson, Shannon, disparity, and relative and absolute variety. Table 2 provides descriptive statistics for JCR data in 2016 of the *Science Citation Index* and *Social Sciences Citation Index* combined.

**Table 2**: Network characteristics of the largest component of the matrix based on JCR 2016.

|  | *JCR 2016* |
|---|---|
| **N of journals (nodes)** | 11,487 |
| **Links** | 3,020,242 |
|  | (11,166 loops) |
| **Total citations** | 50,030,365 |
| **Density** | 0.023 |
| **Average (total) degree** | 525,854 |
| **Cluster coefficient** | 0.221 |
| **Avg. distance** | 2.469 |
| **Maximum distance** | 7 |

As noted, the aggregated citation relations among more than 10k journals provide us with a rich domain containing cited and citing distributions for each of the journals that we can input in Eqs. 1 and 2. However, diversity can be measured in any set of values. The measure is a statistic and therefore dimensionless.

2.2.    Methods

The data is first organized into a citation matrix of 11,487 journals citing one another. RS diversity, Gini, Simpson, and the new diversity index are vector-based and therefore different along the column or row vectors (cited or citing). Our routine available at http://www.leydesdorff.net/software/mode2div/ (see Annex I) operates on column vectors. However, the matrix can easily be transposed within Pajek by following the menus *Network > Create New Network > Transform > Transpose*.

The routine prompts for a .net-file in the Pajek format containing the network data in a 2-mode matrix. (If the matrix is 1-mode, one can use within Pajek: *Network > Create New Network> Transform> 1-Mode to 2-Mode*.) We recommend saving the network in Pajek itself in order to make sure that the format is standardized. The routine also assumes the presence in the same folder of a file coocc.net in Pajek format with a matrix of co-occurrences among vectors in the other (column) direction. One can generate this latter file in Pajek from the initial data file as follows: *Network > 2-Mode Network > 2-Mode to 1-Mode > Rows*. The (dis)similarities (e.g., *cosine* values) are computed among the classes which are used for the comparison. For example, when twenty cities are compared in terms of 654 patent classes one needs the 654 * 654 (dis)similarity matrix for comparing the column vectors of 654 cells representing the twenty cities.

The co-occurrence values can be used as numerators in a large number of (dis)similarity measures (Jones & Furnas, 1987). In this study, we will use (1 – *cosine*) as a measure of the distance in the disparity term [$\sum_{i,j} d_{ij}$]. The cosine is a convenient (non-parametric) measure which varies between 0 and 1, disregards the zeros (Ahlgren, Jarneving, & Rousseau, 2003), and does not assume normality in the distribution. However, other (dis)similarity measures can also be used (e.g., Jaccard, Euclidean distances, etc.).

The classical definition of the Gini coefficient is as follows:

$$G = \frac{\sum_{i=1}^{n}\sum_{j=1}^{n}\left|x_i - x_j\right|}{2n^2 \bar{x}} \qquad (3)$$

where *x* is an observed value, *n* is the number of values observed, and *x bar* is the mean value. If the *x* values are first placed in ascending order, such that each *x* has rank *i*, some of the comparisons above can be avoided:

$$G = \frac{2}{n^2 \bar{x}} \sum_{i=1}^{n} i(x_i - \bar{x}) \qquad (3a)$$

$$G = \frac{\sum_{i=1}^{n}(2i-n-1)x_i}{n \sum_{i=1}^{n} x_i} \qquad (3b)$$

where *x* is an observed value, *n* is the number of values observed, and *i* is the rank of values in ascending order.

The output file div_col.dbf contains (1) RS diversity, (2) "true diversity" which is equal to [1 / (1 – RS Div) ] as derived by Zhang *et al.* (2017, Eq. 6 at p. 1260); (3) the new DIV measure; (4) the Gini coefficient; (5) Simpson Index; (6) Shannon entropy; (7) disparity; and (8) both relative and absolute variety. The routine runs over the lower triangle of the symmetrical (dis)similarity matrix and multiplies the result by two when appropriate. We added betweenness centrality (BC) and journal impact factors (JIF2) of all journals separately.

## 4. Results

*4.2. The full set in the JCR 2016 (n = 11,487)*

Tables 3 and 4 compare the top-25 ranked journals in the citing and cited directions sorted by the different indicators under discussion. Using RS diversity in the citing dimension, we see the unsatisfying result of having the *J Mar Sci Tech-Taiw* as the leading ranked journal in 2016, replacing the *J Chin Inst Eng,* which led this ranking in 2015. The latter journal is now in the second position. This top-25 list is mainly composed of journals with a national identity of less-developed nations. These journals may indeed be transdisciplinary in citing across very different fields of science as a form of local knowledge integration, but this is not the type of "interdisciplinarity" which is valued in the sciences or in the science policy domain (Wagner *et al.*, 2011).

Using the new indicator, Table 4 shows findings where the more obvious candidates for "interdisciplinarity" such as *PLOS ONE, Sci Rep-UK*, etc., are indicated in the *cited* dimension. *Science* and *Nature*, however, are not among the top-25 in the *citing* direction because referencing is very precise and disciplined *within* these journals. This accords with the intuition that articles in these two journals are cited broadly because of their status. In the cited direction, *PNAS* follows in the fifth position after *PLOS ONE* in the first position and *The Lancet* in the fourth.

RS diversity in the cited dimension brings general social science journals to the fore: *Daedalus-US, Qual Inq, Crit Inq,* and *Soc Res* lead this ranking. Perhaps, referencing to these journals is less codified than in the natural or life sciences. Interestingly, ranking the journals on the Gini index brings the natural sciences to the fore; and ranking on the Simpson index foregrounds the major medical journals. For information purposes, these (vector-based) indicators are added to the listings in Tables 3 and 4.

The Spearman rank-order correlations are provided in Annex II. The new diversity indicator correlates much more highly with BC than RS diversity: Spearman's $\rho =.51$ versus .10 in the citing dimension, and .66 versus .41 in the cited one. RS diversity is different from both BC and the new diversity measure; the two measures rank-order correlate only .19 between them in the citing dimension and .35 in the cited. RS diversity is not or negatively correlated to JIF2.

| Journal | BC*1000 | Journal | DIV Citing | Journal | RS Citing | Gini | Simpson |
|---|---|---|---|---|---|---|---|
| PLOS ONE | 149.462 | Springerplus | 0.146 | J Mar Sci Tech-Taiw | 0.942 | Mon Not R Astron Soc | Springerplus |
| Sci Rep-Uk | 51.368 | PLOS ONE | 0.122 | J Chin Inst Eng | 0.941 | Astrophys J | Int J Clin Exp Med |
| P Natl Acad Sci Usa | 45.904 | Biomed Res Int | 0.097 | Teh Vjesn | 0.939 | Astron Astrophys | Biomed Res Int |
| Soc Sci Med | 25.847 | Sci Rep-Uk | 0.093 | Arab J Sci Eng | 0.937 | Phys Rev D | Exp Ther Med |
| Nature | 17.842 | Int J Env Res Pub He | 0.081 | J Cent South Univ | 0.937 | J High Energy Phys | Turk J Med Sci |
| Science | 16.732 | Int J Mol Sci | 0.072 | Dyna-Bilbao | 0.937 | Phys Rev B | Biomed Res-India |
| Int J Env Res Pub He | 16.361 | Peerj | 0.071 | Measurement | 0.931 | Phys Rev A | PLOS ONE |
| Energ Policy | 14.839 | Int J Clin Exp Med | 0.070 | Adv Mech Eng | 0.931 | Org Lett | Int J Mol Sci |
| Psychol Bull | 14.222 | Exp Ther Med | 0.066 | Sci Iran | 0.929 | Phys Rev C | Medicine |
| Sustainability-Basel | 12.142 | Medicine | 0.062 | J Eng Res-Kuwait | 0.928 | J Geophys Res-Space | Curr Pharm Design |
| Nat Commun | 10.837 | Curr Pharm Design | 0.060 | J Fac Eng Archit Gaz | 0.927 | Ieee T Power Electr | Int J Env Res Pub He |
| Am J Public Health | 9.277 | Molecules | 0.057 | Ing Invest | 0.926 | Astron J | Biomed Pharmacother |
| Scientometrics | 8.670 | Med Sci Monitor | 0.055 | Sustainability-Basel | 0.925 | Astrophys J Lett | Drug Des Dev Ther |
| Manage Sci | 7.967 | Eur Rev Med Pharmaco | 0.052 | Sadhana-Acad P Eng S | 0.925 | J Org Chem | Med Sci Monitor |
| Biomed Res Int | 7.357 | Sustainability-Basel | 0.050 | P Est Acad Sci | 0.925 | Phys Rev Lett | Appl Sci-Basel |
| Ecol Econ | 7.288 | Sensors-Basel | 0.050 | Appl Sci-Basel | 0.921 | J Chem Phys | Chinese Med J-Peking |
| Sci Total Environ | 7.137 | Evid-Based Compl Alt | 0.048 | Math Probl Eng | 0.919 | Atmos Chem Phys | Med Hypotheses |
| Front Psychol | 6.595 | Curr Sci India | 0.047 | Qual Quant | 0.915 | Astrophys J Suppl S | Arab J Sci Eng |
| Phys Rev E | 6.258 | Bmj Open | 0.046 | Sci China Technol Sc | 0.915 | Phys Lett B | Iran J Basic Med Sci |
| Sensors-Basel | 6.009 | Front Pharmacol | 0.046 | J Zhejiang Univ-Sc A | 0.913 | J Am Chem Soc | An Acad Bras Cienc |
| Global Environ Chang | 5.694 | Biomed Pharmacother | 0.046 | Maejo Int J Sci Tech | 0.913 | Chem-Eur J | Iran Red Crescent Me |
| Psychol Rev | 5.453 | Cochrane Db Syst Rev | 0.045 | Convergencia | 0.912 | Angew Chem Int Edit | Acta Medica Mediterr |
| Annu Rev Psychol | 5.078 | Med Hypotheses | 0.045 | Rev Estud Soc | 0.911 | Chem Commun | Sains Malays |
| Int J Mol Sci | 5.030 | Curr Med Chem | 0.045 | Ksce J Civ Eng | 0.909 | J Cosmol Astropart P | Braz Arch Biol Techn |
| Springerplus | 4.769 | Mol Med Rep | 0.045 | Ieee Lat Am T | 0.909 | Eur Phys J C | Iran J Public Health |

**Table 3**: Top 25 journals in terms of betweenness centrality (BC) and various diversity measures in the citing direction.

| Journal | Bc * 1000 | Journal | Div Cited | Journal | Rs Cited | Gini | Simpson |
|---|---|---|---|---|---|---|---|
| PLOS ONE | 149.462 | PLOS ONE | 0.142 | Daedalus-Us | 0.939 | Mon Not R Astron Soc | New Engl J Med |
| Sci Rep-Uk | 51.368 | Science | 0.125 | Qual Inq | 0.936 | Astrophys J | Jama-J Am Med Assoc |
| P Natl Acad Sci Usa | 45.904 | Nature | 0.124 | Crit Inquiry | 0.927 | Astron Astrophys | Lancet |
| Soc Sci Med | 25.847 | Lancet | 0.121 | Soc Res | 0.927 | Phys Rev D | Ann Intern Med |
| Nature | 17.842 | P Natl Acad Sci Usa | 0.114 | Comput J | 0.926 | Astron J | Ann Ny Acad Sci |
| Science | 16.732 | Ann Ny Acad Sci | 0.113 | Qual Quant | 0.922 | J High Energy Phys | Am J Med |
| Int J Env Res Pub He | 16.361 | New Engl J Med | 0.108 | Field Method | 0.922 | Astrophys J Lett | Drugs |
| Energ Policy | 14.839 | Jama-J Am Med Assoc | 0.103 | New Left Rev | 0.917 | Astrophys J Suppl S | Science |
| Psychol Bull | 14.222 | Bmj-Brit Med J | 0.096 | Qual Res Psychol | 0.917 | J Geophys Res-Space | Nature |
| Sustainability-Basel | 12.142 | Biomed Res Int | 0.093 | Am Behav Sci | 0.915 | Phys Rev Lett | Anal Biochem |
| Nat Commun | 10.837 | Ann Intern Med | 0.083 | P Ieee | 0.911 | Nucl Phys B | Psychol Bull |
| Am J Public Health | 9.277 | Cochrane Db Syst Rev | 0.081 | Risk Anal | 0.908 | Phys Lett B | P Ieee |
| Scientometrics | 8.670 | AM J MED | 0.077 | STATA J | 0.908 | PHYS REV A | LIFE SCI |
| Manage Sci | 7.967 | Psychol Bull | 0.074 | Qual Res | 0.908 | Org Lett | Philos T R Soc A |
| Biomed Res Int | 7.357 | Biochem Bioph Res Co | 0.073 | Psychometrika | 0.907 | J Am Chem Soc | Pharmacol Rev |
| Ecol Econ | 7.288 | Int J Mol Sci | 0.071 | Commun Acm | 0.906 | Publ Astron Soc Pac | Pharmacol Therapeut |
| Sci Total Environ | 7.137 | Sci Rep-Uk | 0.069 | Sociol Method Res | 0.905 | Atmos Chem Phys | Am Psychol |
| Front Psychol | 6.595 | Mayo Clin Proc | 0.068 | Eur J Soc Theory | 0.905 | Angew Chem Int Edit | Biomed Res Int |
| Phys Rev E | 6.258 | J Clin Invest | 0.068 | Signs | 0.904 | Phys Rev B | Adv Exp Med Biol |
| Sensors-Basel | 6.009 | Am J Epidemiol | 0.064 | Econ Soc | 0.903 | Phys Rev C | J Econ Perspect |
| Global Environ Chang | 5.694 | Clin Chem | 0.064 | Sociology | 0.902 | Icarus | Can Med Assoc J |
| Psychol Rev | 5.453 | Faseb J | 0.064 | Hist Hum Sci | 0.902 | Annu Rev Astron Astr | J Intern Med |
| Annu Rev Psychol | 5.078 | J Biol Chem | 0.064 | Educ Psychol Meas | 0.901 | J Org Chem | Mayo Clin Proc |
| Int J Mol Sci | 5.030 | Biometrics | 0.063 | Siam Rev | 0.901 | Ieee T Power Electr | Am J Med Sci |
| Springerplus | 4.769 | Can Med Assoc J | 0.063 | Ieee Spectrum | 0.901 | Mitochondr Dna | Postgrad Med J |

**Table 4**: Top 25 journals in terms of betweenness centrality (BC) and various diversity measures in the cited direction.

The rank-order correlation of the two indicators for the cited patterns of the 11,467 journals[11] is only .35 ($p<.01$), and as low as .19 ($p<.01$) in the citing direction. This suggests that the two measures indicate different things. The various correlations are provided in Annex II. However, we will take this question one step further (below) by factor analyzing the indices in order to answer the question of whether they represent different dimensions or essentially the same ones.

|  | Citing | | Cited | |
| --- | --- | --- | --- | --- |
|  | RS | DIV | RS | DIV |
| GINI | -.362 | .217 | -.091 | .285 |
|  | -.394 | .292 | -.114 | .359 |
| SIMPSON | .276 | .434 | .537 | .419 |
|  | .175 | .774 | .387 | .845 |
| SHANNON | .217 | .802 | .485 | .790 |
|  | .145 | .960 | .395 | .974 |
| VARIETY | -0.006 | .895 | .191 | .932 |
|  | -.032 | .922 | .313 | .958 |
| DISPARITY | .703 | .062 | .695 | .229 |
|  | .687 | .017 | .680 | .262 |

**Table 5**: Correlation of various components of diversity with the two diversity measures, both citing and cited. All correlations are significant at the 1%-level; top-lines provide Pearson correlations, bottom-lines Spearman's rank-order correlations.

---

[11] Of the 11,487 journals, twenty are not cited (Table 7b) and (11,487 – 11,298 =) 489 are not registered in terms of references.

Table 5 shows correlations between the two diversity indicators under discussion (RS and DIV) and the various components discussed above. RS diversity correlates highest with disparity, while DIV correlates considerably more with the other components such as the Gini index, Simpson, Shannon, and relative variety. In the case of RS, disparity is more prominent in the result of the multiplication (Eq. 1 above) because it is multiplied by only a single other factor (the Simpson index), while it is multiplied by two other components in the case of DIV. Thus, the synthesis into "dual concept" diversity reduces the influence of variety on RS diversity.

As noted, Stirling (1998, p. 57) concluded that there are good reasons to prefer the Shannon measure above the Simpson Index *if* one wishes to measure the two concepts in a single operationalization as a "dual concept." The respective differences in terms of the correlations with Shannon entropy illustrate our point, since DIV correlates much high with Shannon entropy than RS.

*4.1.    Eighty-five journals in the Library and Information Sciences* (2016)

Eighty-five journals were assigned to the WC labeled "information science & library science" in JCR 2016. We study the asymmetrical citation matrix among these 85 journals in both the cited and citing directions. Table 6 provides the 25 highest ranking-journals in the various dimensions of "citing"; Table 7 the corresponding values in the cited direction.

|  | BC *100 | Journal | DIV citing | Journal | RS Citing |
|---|---|---|---|---|---|
| J Assoc Inf Sci Tech | 73.114 | J Assoc Inf Sci Tech | 0.193 | Inform Soc-Estud | 0.708 |
| Scientometrics | 61.592 | Inform Dev | 0.180 | Investig Bibliotecol | 0.684 |
| Investig Bibliotecol | 44.759 | Aslib J Inform Manag | 0.175 | Aslib J Inform Manag | 0.667 |
| J Inf Sci | 39.379 | Libr Hi Tech | 0.175 | Rev Esp Doc Cient | 0.666 |
| Inform Soc-Estud | 37.509 | Electron Libr | 0.163 | Transinformacao | 0.661 |
| Inform Dev | 34.427 | Inform Res | 0.148 | Afr J Libr Arch Info | 0.655 |
| J Acad Libr | 33.019 | J Libr Inf Sci | 0.145 | Electron Libr | 0.654 |
| Libr Hi Tech | 28.734 | J Inf Sci | 0.139 | Inform Technol Dev | 0.643 |
| Mis Quart | 22.785 | Can J Inform Lib Sci | 0.135 | Libr Hi Tech | 0.641 |
| Inform Manage-Amster | 22.773 | Program-Electron Lib | 0.129 | Can J Inform Lib Sci | 0.633 |
| Libri | 22.567 | Libr Inform Sci Res | 0.121 | Libri | 0.633 |
| Coll Res Libr | 20.215 | Online Inform Rev | 0.118 | Prof Inform | 0.625 |
| J Knowl Manag | 19.651 | Investig Bibliotecol | 0.118 | Online Inform Rev | 0.624 |
| Prof Inform | 19.512 | Libri | 0.113 | J Assoc Inf Sci Tech | 0.606 |
| Libr Quart | 19.441 | Inform Technol Dev | 0.112 | Program-Electron Lib | 0.605 |
| Electron Libr | 14.239 | Int J Inform Manage | 0.111 | Interlend Doc Supply | 0.596 |
| Inform Process Manag | 13.721 | J Doc | 0.108 | J Libr Inf Sci | 0.595 |
| Interlend Doc Supply | 13.605 | J Acad Libr | 0.107 | Inform Dev | 0.585 |
| Online Inform Rev | 12.195 | Telemat Inform | 0.103 | Telemat Inform | 0.580 |
| Telemat Inform | 11.761 | Malays J Libr Inf Sc | 0.102 | Serials Rev | 0.577 |
| J Health Commun | 11.626 | Rev Esp Doc Cient | 0.095 | Malays J Libr Inf Sc | 0.559 |
| Inform Technol Peopl | 11.583 | Prof Inform | 0.093 | Knowl Organ | 0.541 |
| Gov Inform Q | 11.513 | J Organ End User Com | 0.092 | Aust Acad Res Libr | 0.539 |
| Ref User Serv Q | 11.474 | Inform Soc-Estud | 0.090 | Gov Inform Q | 0.533 |
| Restaurator | 11.331 | J Inf Technol | 0.080 | Inform Res | 0.530 |

**Table 6**: Top 25 LIS journals in terms of betweenness centrality (BC) and various diversity measures in the citing direction.

| Journal | DIV cited | Journal | RS Cited |
|---|---|---|---|
| Scientometrics | 0.155 | Inform Organ-Uk | 0.696 |
| J Am Med Inform Assn | 0.149 | Libr Quart | 0.678 |
| J Informetr | 0.144 | Libr Inform Sci Res | 0.669 |
| J Glob Inf Tech Man | 0.143 | Libr Inform Sc | 0.645 |
| J Glob Inf Manag | 0.138 | J Inf Sci | 0.635 |
| Portal-Libr Acad | 0.135 | Investig Bibliotecol | 0.632 |
| J Scholarly Publ | 0.133 | J Knowl Manag | 0.632 |
| Ethics Inf Technol | 0.133 | Libr Resour Tech Ser | 0.631 |
| Knowl Man Res Pract | 0.125 | Soc Sci Comput Rev | 0.611 |
| Int J Geogr Inf Sci | 0.120 | Online Inform Rev | 0.608 |
| Telecommun Policy | 0.119 | Program-Electron Lib | 0.585 |
| Qual Health Res | 0.118 | J Health Commun | 0.582 |
| J Med Libr Assoc | 0.118 | Restaurator | 0.572 |
| Electron Libr | 0.117 | Aust Libr J | 0.570 |
| J Assoc Inf Sci Tech | 0.116 | Gov Inform Q | 0.567 |
| Inf Tarsad | 0.115 | Libr Hi Tech | 0.567 |
| Mis Q Exec | 0.111 | Learn Publ | 0.563 |
| Inform Process Manag | 0.107 | Can J Inform Lib Sci | 0.556 |
| Inform Technol Peopl | 0.106 | Inform Technol Libr | 0.544 |
| J Acad Libr | 0.106 | Inform Process Manag | 0.543 |
| Interlend Doc Supply | 0.105 | J Acad Libr | 0.525 |
| Coll Res Libr | 0.105 | Afr J Libr Arch Info | 0.524 |
| Serials Rev | 0.104 | Econtent | 0.508 |
| Int J Comp-Supp Coll | 0.103 | Aust Acad Res Libr | 0.499 |
| Inform Technol Manag | 0.103 | Inform Dev | 0.489 |

**Table 7**: Top 25 LIS journals in terms of betweenness centrality (BC) and various diversity measures in the cited direction.

In the citing direction, the new indicator (rank-order) correlates with BC with $\rho = .59$ ($p<.01$), whereas RS diversity correlates with BC with $\rho = .35$ ($p<.01$). The two indicators—DIV and RS—correlate .75 ($p<.01$). In the cited direction (Tables 7 and A2b), the Spearman rank-order correlations with BC are .58 ($p<.01$) and .27 ($p<.05$), for DIV and RS respectively. In sum, the new indicator correlates with BC considerably more than RS diversity, both in the cited and the citing dimension. See Annex III for Tables with the correlations. We added to Table A3a and A3b in Annex III a column and row providing the correlations with JIF2 for the orientation of the reader. For this smaller group of 85 journals, both RS and DIV correlate not significantly with JIF2 in both the cited and citing dimensions.

Interdisciplinarity in the citing dimension indicates knowledge integration and one expects more marginal journals to take this role, whereas larger and more leading journals can be expected to have a role in interdisciplinary knowledge diffusion (cited). This difference is reflected in the values for DIV in Table 7, but not for RS. *Scientometrics* and *JASIST* [12] lead the ranking in terms of DIV in the cited direction, but are not among the top-25 journals when using RS. *Scientometrics* has the 69$^{th}$ position on this list of 85 journals when ranked using RS. As noted, this was the 70$^{st}$ position in our previous study using 2015 data; this finding triggered our worries about using RS for measuring interdisciplinarity.

## 5. Factor Analysis

Factor analysis allows us to test whether the indicators (RS, DIV, and BC) measure the same or different dimensions (of diversity). We used principal component analysis as extraction method and rotated using varimax in SPSS. The values of the Gini coefficient and the Simpson index are components of the diversity measures under discussion; inclusion of these variables into the factor analysis would therefore be redundant. However, we added JIF2 for the orientation of the reader as a kind of benchmark.

---

[12] The full journal name of *JASIST* was changed in 2014 from *Journal of the American Society for Information Science and Technology* into *Journal of the Association of Information Science and Technology*.) The journal can therefore no longer cite under its old name, but it is cited mainly under this name.

Two factors (components) in the analysis have eigenvalues higher than 1; the two factors explain 65.1% of the variance in the data. Table 8 shows the results of the factor analysis: the factor loadings of the different diversity measures on the two components. In the interpretation of the results, we focus on factor loadings with values greater than 0.5 (boldfaced in Table 8).

**Rotated Component Matrix[a]**

|  | Component 1 | Component 2 |
|---|---|---|
| DIV_cited | **.896** | .061 |
| DIV_citing | **.693** | .400 |
| JIF2 | **.647** | -.329 |
| BC | **.502** | .104 |
| RS_citing | -.107 | **.851** |
| RS_cited | .253 | **.702** |

Extraction Method: Principal Component Analysis.
Rotation Method: Varimax with Kaiser Normalization.
a. Rotation converged in 3 iterations.

**Table 8**: Rotated component matrix of relevant diversity indicators of 11,467 journals in the JCR 2016 for two dimensions. (Varimax rotated with Kaiser Normalization in SPSS.)

The new DIV indicators (cited and citing) load on one factor (component)—together with BC (and JIF2). The other factor (dimension) is determined by both RS indicators (cited and citing). DIV in the *cited* direction (interdisciplinary diffusion) has the highest loading on the first factor and is completely uncoupled from Factor 2. The latter factor couples to interdisciplinary knowledge integration with highest factor loading for RS in the *citing* direction.

## 6. Range

Figure 2 shows the ranges of RS diversity and the new diversity measure across the 11,487 journals on a log-log scale. Of these 11,487 journals, 10,264 (89.3%) have RS-diversity values above 0.5. In other words, most journals are indicated as diverse. However, the lower values and the larger spread of the new diversity indicator is a consequence of multiplying three terms between zero and one, while only two terms (< 1) are multiplied in the case of RS diversity.[13] However, the much larger range allows for more refined measurement

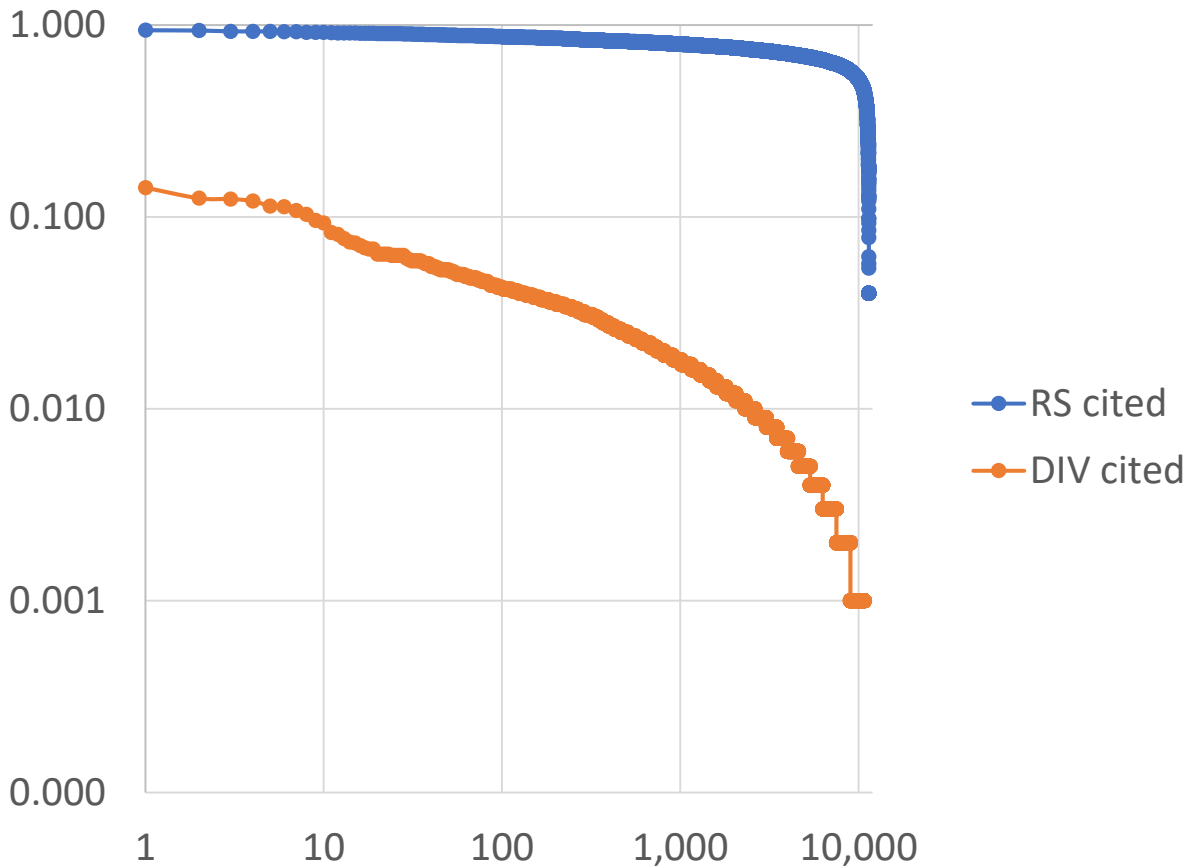

**Figure 2**: Ranges of RS diversity and the new diversity measure across 11,487 journals covered by the JCR 2016.

---

[13] For example: 0.6 * 0.6 = 0.36, but 0.6 * 0.6 * 0.6 = 0.216.

## 7. Summary and conclusions

We asked whether the measurement of interdisciplinarity can be improved by using a new measure of diversity (DIV), when compared with RS. We have shown that the three components of diversity (variety, balance, and disparity) can be measured independently and then combined, creating a more informed result. We operationalized these three components independently as follows:

1. "balance" is operationalized by the Gini coefficient;
2. "variety" is operationalized as "relative variety," that is, the number of classes in use divided by the number of classes available for use;
3. "disparity" can be operationalized using one distance measure or another as in the case of RS diversity; in this study, we used (1 – *cosine*).

The three components are all bounded between zero and one; diversity is measured by multiplying the three values for each element. The resulting indicator is then necessarily bound between zero and one.

In comparison to RS diversity, the new indicator (DIV) has the following advantages:

1. It is no longer based on "dual-concept" diversity, but on the independent operationalization of the three components of diversity: balance, variety, and disparity.

2. It is monotonic (Rousseau, *personal communication*, 18 June 2018): diversity increases for each of the three components when the other two remain the same.
3. "Balance" is operationalized as the Gini coefficient; including this indicator as a component provides greater specificity to the resulting indicator of diversity.
4. The empirical results of the measurement are less puzzling and counterintuitive that one doubts the values of the indicator because of possible flaws in its construction.
5. The new indictor correlates with betweenness centrality significantly more than RS diversity.

Comparing the measurements of two different indicators, one can always expect the results to be different. At best, they may point in the same direction (Hicks *et al*., 2015). Since there is no ground-truth of "interdisciplinarity," there are no obvious criteria to choose one indicator over another on the basis of empirical results unless the results show obvious (in)validity. This problem has been pointed out before, for instance by Stirling (2007). Furthermore, "interdisciplinarity" is based on the underlying concept of "disciplines" which are social constructs developed to allocate the privileges and responsibilities of expertise and the allocation of resources (Wagner, forthcoming). The boundaries among disciplines, however, are fluid.

In our case, the initial reason to deconstruct RS diversity was the puzzles it continued to pose when measuring diversity as an indicator of interdisciplinarity. These results were often incomprehensible and sometimes counter-intuitive. Via a series of communications, we came to the conclusion that RS is flawed as a measure of diversity because of the *method* of combining the three components of variety, balance, and disparity. In our opinion, the problem is the *ex ante* combination of variety and balance as a "dual-concept" indicator; there is no theoretical reason nor practical need for this shortcut.[14]

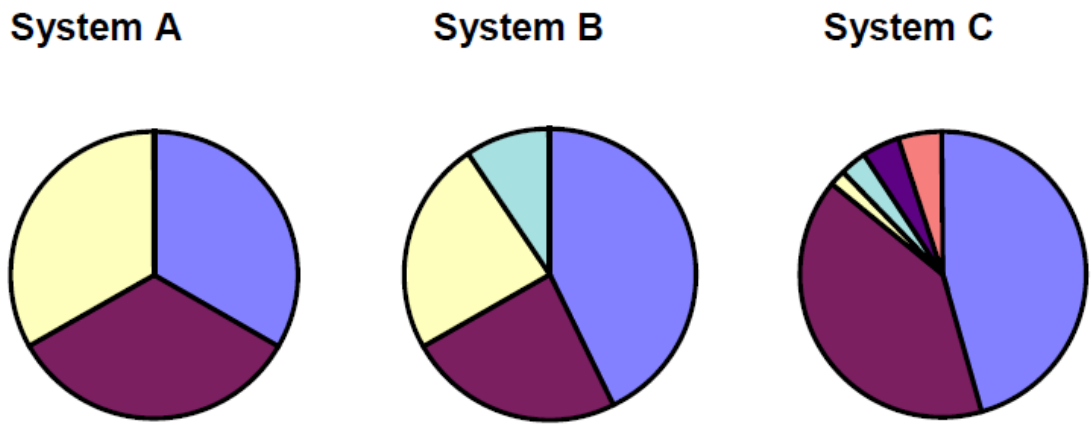

**Figure 7**: The relative priority assigned to variety and balance in dual concept diversity. Source: Stirling (1998, at p. 49).

Stirling (1998) used Figure 7 to show the dilemma when combining the two "subordinate properties" of variety and balance into a single "dual concept": "Where variety is held to be the most important property, System C might reasonably be held to be most (dual concept) diverse. Where a greater priority is attached to the evenness in the balance between options, System A might be ranked highest. In addition, there are a multitude of possible intermediate possibilities, such as System B" (Stirling, 1998, p. 48).

---

[14] Stirling (2007) was probably not aware of the possibility of using the Gini coefficient as an indicator of balance. There are no references to Rousseau's work, and the Gini is classified in Table 1 (at p. 709) as a "dual-concept" measure of variety and balance.

Rousseau (2018a) added that RS is (*i*) not monotonic despite its aspiration to fulfill this requirement, and (*ii*) that "balance" is not even covered by RS diversity despite its crucial role in Stirling's (2007) theoretical argument. Nijssen, Rousseau, & Van Hecke (1998), however, have proven that "balance" can be indicated by the Gini coefficient. The Gini coefficient is conveniently bounded between zero and one, and relative variety can also be defined between zero and one (as $n_c / N$). Thus, the reasoning behind Stirling (2007) can be conserved, but in the case of DIV the operationalization is changed, expanded, and made more specific.

**Acknowledgment**

We thank Lin Zhang and Ronald Rousseau for comments on a previous version of the manuscript.

**Annex I**

The program mode2div.exe (at https://leydesdorff.github.io/diversity_measurement/ or http://www.leydesdorff.net/software/mode2div) computes diversity measures along the column vectors of a 2-mode matrix saved in the .net format of Pajek. (Use preferentially Pajek itself for the saving.) The program and the data file have to be in the same folder. One is first prompted for the name of the file with the data.

If the matrix is 1-mode, use the following instruction in Pajek: *Network > Create New Network> Transform> 1-Mode to 2-Mode.* Save the file as a 2-mode matrix.

Additionally, one needs a file <coocc.net> made in Pajek. One can generate this file using the 2-mode file in Pajek as follows: *Network > 2-Mode Network > 2-Mode to 1-Mode > Rows.* Save the resulting file as coocc.net. (Obligatory file name; the file needs to be in the same folder.) Note that the co-occurrences have to be taken along the rows, when the column vectors are compared. Unlike the Gini or Simpson, RS and DIV are not measures of the diversity at the vector level, but at the level of the matrix (including disparity).

Output is the file div_col.dbf containing the various diversity indicators along the column dimension for all units of analysis (in the rows). (One can transpose the data matrix in Pajek: *Network > Create New Network > Transform > Transpose*. Thereafter, coocc.net has also to be replaced.) Note that previous versions of the output file ("div_col.dbf") are overwritten during subsequent runs.

**Annex II:**

Spearman rank-order correlations between the proposed measure of diversity (DIV), RS diversity, Betweenness Centrality (BC), and the Journal Impact Factor (JIF2) for 11,487 journals included in the JCR 2015.

**Correlations**

|     |                 | RS       | DIV     | BC      | JIF2     |
|-----|-----------------|----------|---------|---------|----------|
| RS  | Correlation     | 1        | .185**  | .096**  | -.293**  |
|     | Sig. (2-tailed) |          | .000    | .007    | .000     |
|     | N               | 11288    | 11115   | 11237   | 11254    |
| DIV | Correlation     | .173**   | 1       | .512**  | .409**   |
|     | Sig. (2-tailed) | .000     |         | .000    | .000     |
|     | N               | 11115    | 11115   | 11073   | 11081    |
| BC  | Correlation     | .025**   | .320**  | 1       | .489**   |
|     | Sig. (2-tailed) | .007     | .000    |         | .000     |
|     | N               | 11237    | 11073   | 11241   | 11217    |
| JIF | Correlation     | -.213**  | .150**  | .068**  | 1        |
|     | Sig. (2-tailed) | .000     | .000    | .000    |          |
|     | N               | 11254    | 11081   | 11217   | 11448    |

**. Correlation is significant at the 0.01 level (2-tailed).

**Table A2a**: Correlations among citing vectors of journals in the JCR 2016.
Upper triangle Spearman's ρ; lower triangle: Pearson's *r*.

**Correlations**

|     |                 | RS      | DIV     | BC      | JIF     |
|-----|-----------------|---------|---------|---------|---------|
| RS  | Correlation     | 1       | .349**  | .408**  | .084**  |
|     | Sig. (2-tailed) |         | .000    | .000    | .040    |
|     | N               | 11421   | 10705   | 11233   | 11395   |
| DIV | Correlation     | .256**  | 1       | .658**  | .609**  |
|     | Sig. (2-tailed) | .000    |         | .000    | .000    |
|     | N               | 10705   | 10705   | 10544   | 10695   |
| BC  | Correlation     | .055**  | .305**  | 1       | .489**  |
|     | Sig. (2-tailed) | .000    | .000    |         | .000    |
|     | N               | 11233   | 10544   | 11241   | 11217   |
| JIF | Correlation     | .019*   | .448**  | .068**  | 1       |
|     | Sig. (2-tailed) | .040    | .000    | .000    |         |
|     | N               | 11395   | 10695   | 11217   | 11448   |

**. Correlation is significant at the 0.01 level (2-tailed).
*. Correlation is significant at the 0.05 level (2-tailed).

**Table A2b**: Correlations among cited vectors of journals in the JCR 2016.
Upper triangle: Spearman's ρ; lower triangle: Pearson's *r*.

**Annex III:**

Spearman rank-order correlations between the proposed measure of diversity (DIV), RS diversity, and Betweenness Centrality (BC), and the Journal Impact Factor (JIF2) for 85 LIS journals included in the JCR 2016.

**Correlations**

|  |  | RS | DIV | BC | JIF2 |
|---|---|---|---|---|---|
| RS | Correlation | 1 | .749** | .354** | -.482** |
|  | Sig. (2-tailed) |  | .000 | .001 | .000 |
|  | N | 79 | 79 | 78 | 79 |
| DIV | Pearson Correlation | .708** | 1 | .582** | -.187 |
|  | Sig. (2-tailed) | .000 |  | .000 | .100 |
|  | N | 79 | 79 | 78 | 79 |
| BC | Correlation | .274* | .510** | 1 | .038 |
|  | Sig. (2-tailed) | .015 | .000 |  | .740 |
|  | N | 78 | 78 | 79 | 79 |
| JIF2 | Correlation | -.426** | -.153 | .032 | 1 |
|  | Sig. (2-tailed) | .000 | .179 | .780 |  |
|  | N | 79 | 79 | 79 | 85 |

**. Correlation is significant at the 0.01 level (2-tailed).
*. Correlation is significant at the 0.05 level (2-tailed).

**Table A3a**: Correlations among citing vectors of 85 LIS journals in the JCR 2016. Upper triangle Spearman's ρ; lower triangle: Pearson's *r*.

**Correlations**

|  |  | RS | DIV | BC | JIF2 |
|---|---|---|---|---|---|
| RS | Correlation | 1 | .682** | .274* | -.201 |
|  | Sig. (2-tailed) |  | .000 | .016 | .073 |
|  | N | 81 | 81 | 77 | 81 |
| DIV | Correlation | .651** | 1 | .582** | .004 |
|  | Sig. (2-tailed) | .000 |  | .000 | .695 |
|  | N | 81 | 81 | 77 | 81 |
| BC | Correlation | .207 | .521** | 1 | .038 |
|  | Sig. (2-tailed) | .070 | .000 |  | .740 |
|  | N | 77 | 77 | 79 | 79 |
| JIF2 | Correlation | -.183 | .092 | .032 | 1 |
|  | Sig. (2-tailed) | .103 | .414 | .780 |  |
|  | N | 81 | 81 | 79 | 85 |

**. Correlation is significant at the 0.01 level (2-tailed).
*. Correlation is significant at the 0.05 level (2-tailed).

**Table A3b**: Correlations among cited vectors of 85 LIS journals in the JCR 2016. Upper triangle Spearman's ρ; lower triangle: Pearson's *r*.